\documentclass[
  aps,twocolumn,showpacs,preprintnumbers,nofootinbib,prd,
  superscriptaddress,10pt
]{revtex4-1}

\usepackage{graphicx,amssymb,amsmath,amsthm,amsfonts,epsfig,epsf,fixmath}
\usepackage{bm}
\usepackage{dcolumn}
\usepackage{latexsym}
\usepackage{rotating}
\usepackage{longtable}
\usepackage{enumerate}
\usepackage{tensor,multirow}
\usepackage{url}
\usepackage{comment}
\usepackage[linktocpage]{hyperref}
\usepackage[usenames]{color}
\usepackage{epstopdf}
\usepackage{tikz,tikz-3dplot}
\usepackage{aas_macros}
\usepackage{ulem}
\usepackage{slashed}
\usepackage[scr=boondox]{mathalfa}
\usepackage[scr=boondox]{mathalfa}
\usepackage{slashed}
\usepackage{ulem}
\usepackage{tikz,tikz-3dplot}
\usepackage[dvipsnames]{xcolor}

\DeclareFontFamily{U}{min}{}
\DeclareFontShape{U}{min}{m}{n}{<-> udmj30}{}

\newcommand{\GSSI}{Gran Sasso Science Institute (GSSI), I-67100 L'Aquila, Italy}
\newcommand{\GranSasso}{INFN, Laboratori Nazionali del Gran Sasso, I-67100 Assergi, Italy}

\begin{document}

\title{Impact of a third body on binary neutron star tidal interactions}

\author{Meet Khatri}
 \affiliation{Birla Institute of Technology and Science Pilani, Rajasthan, 333031, India}
 \affiliation{Raman Research Institute, Bengaluru, 560080, India}

\author{Ankur Renduchintala}
 \affiliation{Birla Institute of Technology and Science Pilani, Rajasthan, 333031, India}
 \affiliation{International Centre for Theoretical Sciences (ICTS-TIFR), Bengaluru, Karnataka, 560089, India}

\author{Sayak Datta}
 \email{sayak.datta@gssi.it}
 \affiliation{\GSSI}
 \affiliation{\GranSasso}
 
\author{Sajal Mukherjee}
 \email{sajal.mukherjee@pilani.bits-pilani.ac.in}
 \affiliation{Birla Institute of Technology and Science Pilani, Rajasthan, 333031, India}

 \date{\today}

\begin{abstract}
To model a compact binary coalescence, it is commonly assumed that the binary evolves in isolation. In the present work, we relax this assumption by introducing a distant third body and investigate how the third body would affect the binary evolution. We treat this problem perturbatively and study tidal interaction in the binary due to the third body's presence. We introduce appropriate modifications to the equations governing the orbital motions and the evolution equations of the binary component's quadrupole moment. We compute the radiated energy and accumulated dephasing for the binary. Our findings indicate that for extreme-mass-ratio inspirals, these effects would not be significant. However, for binaries such as intermediate-mass-ratio inspirals, these effects are more relevant, and in some scenarios the accumulated dephasing can reach up to $\mathcal{O}(1)$ radians.
\end{abstract}
\maketitle
\section{Introduction}
With the proposed gravitational wave (GW) detectors to operate in the near future, GW astronomy is entering a new phase \cite{Evans:2021gyd, Abac:2025saz, LISA:2024hlh, TianQin:2015yph}. Upcoming detectors are expected to significantly expand the observable GW source population. \cite{Bailes:2021}. In addition, we may be able to observe and understand the perturbative effects, which are generally weak in nature \cite{LISA:2022yao}. These perturbative effects arise from diverse physical mechanisms, making them challenging to model. Together, these perturbative effects are called \textit{environmental effects} and have become an active area of research in recent years \cite{LISAConsortiumWaveformWorkingGroup:2023arg}. Here we focus on one such effect, which is caused by a distant third body.

The presence of a third body can be incorporated in several ways. 
One such possibility is to include the effect of a tertiary companion in an extreme mass-ratio inspiral (EMRI). 
This scenario has been investigated primarily in the context of tidal resonances~\cite{Bonga:2019ycj, Gupta:2021cno, Gupta:2022fbe} and also the impact on the secular dynamics \cite{Camilloni:2023rra, Grilli:2024fds}. 

Another possibility arises when the secondary, instead of being a single stellar-mass compact object, is itself a stellar-mass binary~\cite{Chen:2018axp, Santos:2025ass} (also see \cite{Peissker:2024ade}). 
For example, a binary neutron star (BNS) system with component masses $m_1$ and $m_2$, each of order $\sim M_{\odot}$, orbiting one another and emitting GWs. If an additional object is placed in the vicinity of this BNS, the configuration constitutes a classical three-body system.

EMRIs, involving a primary of mass $\sim 10^{5}-10^7 M_{\odot}$ and a companion of mass $\sim 1-10^2 M_{\odot}$, are observable over $\mathcal{O}(10^4-10^5)$ orbits \cite{Berry:2019wgg}. The secondary evolves within a few gravitational radii of the primary before plunging, emitting millihertz GWs well within the sensitivity of LISA~\cite{Audley:2017drz} and TianQin \cite{TianQin:2015yph}. On the other hand, Intermediate Mass Black Holes (IMBHs; $10^{2}-10^{4}  M_{\odot}$) can form Intermediate Mass Ratio Inspirals (IMRIs) with stellar or supermassive black holes, having mass ratios $q \sim 10^{-4}\!-\!10^{-2}$ \cite{Klein:2015hvg, Volonteri:2020wkx}. IMRIs exhibit shorter inspirals and less variability than EMRIs \cite{Arca-Sedda:2020lso}, emitting across \( 10^{-3}\!-\!10 \) Hz, making them multi-band sources observable by mHz \cite{LISA:2024hlh,2025arXiv250220138L}, decihertz \cite{Ajith:2024mie}, and 3G detectors \cite{Abac:2025saz,2023arXiv230613745E}. Here we focus on these systems while replacing the secondary with a stellar-mass binary.

\begin{figure}[htp]
\centering
\includegraphics[scale=.43]{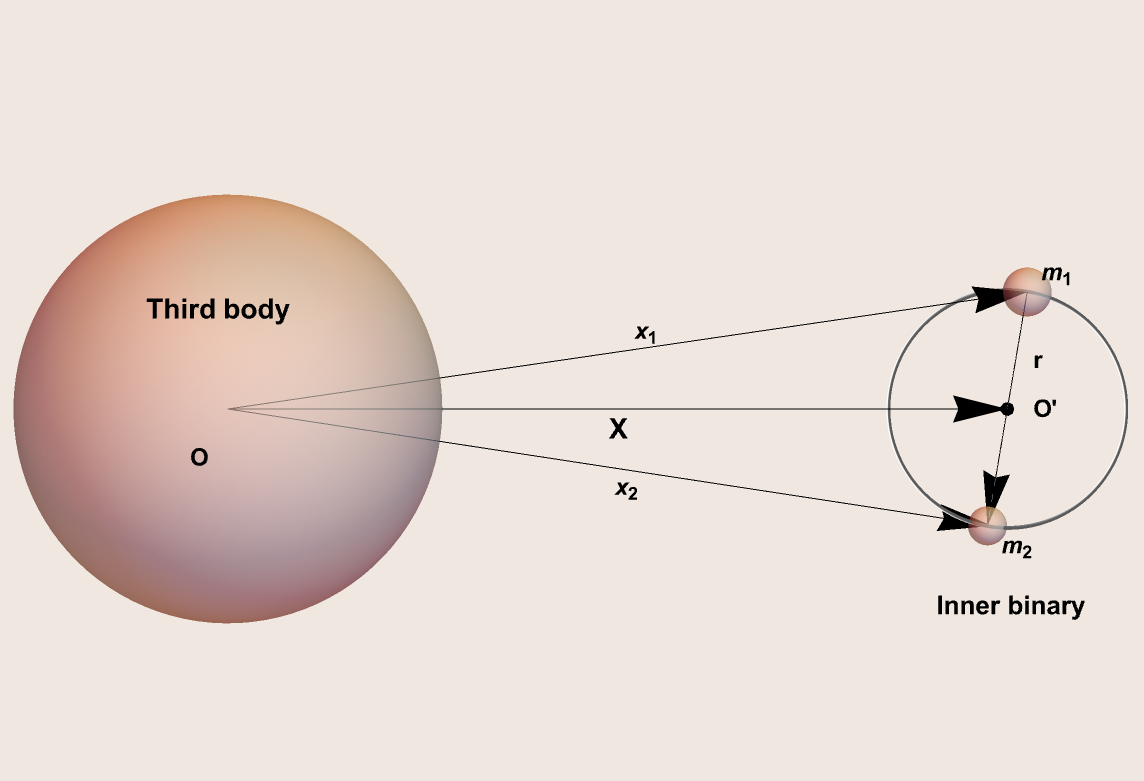}
\caption{A schematic figure of the system under consideration from a top-down view. The inner binary with component masses and radii $m_i$ and $\boldsymbol{x}_i$ has its center of mass (COM) at $O'$. The center of the third body (larger sphere) is at $O$, hence it is the origin of the chosen coordinate system; $\boldsymbol{r}$ is the relative separation of the components of the inner binary; $\boldsymbol{X}$ is the separation of the outer binary comprising the third body and the COM of the inner binary.}
\label{fig_01}
\end{figure}

Specifically, we consider scenarios in which the third body is massive ($\sim 10^2- 10^6 M_{\odot}$) and located at a large distance, covering the full range from IMRIs to EMRIs.
These types of systems are called binary intermediate/extreme mass ratio inspirals or b-IMRIs/b-EMRIs in short \cite{Chen:2018axp}. In Fig. \ref{fig_01}, we pictorially demonstrate one such system composed of an inner binary and a massive third body. 

The possibility of formation of such systems in the realistic astrophysical realm has motivated many studies in recent years \cite{Camilloni:2023xvf, Meiron_2017, Han_2019, Yin:2024nyz}. Compact binaries can form in nuclear star clusters or within accretion disks, and if they remain bound to the super massive black hole (SMBH) without being tidally disrupted, they constitute a stable triple configuration \cite{Antonini:2012ad, Michaely:2014dra, 2017MNRAS.464..946S}. Secular three-body dynamics, such as Kozai-Lidov oscillations, can periodically excite the eccentricity of the inner binary, driving it toward merger within the sphere of influence of the third body \cite{Antonini:2012ad, Antonini:2016gqe, Hoang:2017fvh}. This pathway applies not only to binaries of black holes but also to BNS systems. However, since neutron stars are less common in galactic nuclei due to mass segregation, the possibility of such binaries could be rare \cite{Freitag:2006qf}. A BNS orbiting an SMBH can survive tidal forces provided the binary separation is sufficiently tight and/or placed sufficiently far, thus opening the possibility of b-EMRIs containing neutron stars. If nature allows for the existence of boson stars \cite{Visinelli:2021uve} or exotic compact objects \cite{Cardoso:2019rvt}, they can also become constituents of such systems.

b-IMRIs share the same hierarchical triple structure but with the tertiary mass reduced to the IMBH scale \cite{Chen:2018axp, Fragione:2018nnl, Fragione:2019dtr}.
In both b-EMRIs and b-IMRIs, the eventual coalescence of the inner compact binary may occur while the system remains bound to the central black hole. Such events produce unique signatures~\cite{Cardoso:2021vjq, Santos:2025ass}. Unique resonance features can also arise \cite{Cocco:2025adu, Cocco:2025udb}. These signatures not only enrich the GW landscape but also provide powerful probes of strong-field dynamics in realistic astrophysical environments.

Three-body systems have been extensively studied for decades~\cite{goldstein2011classical,valtonen2006three,1984A&A...141..232S, Michaely:2014dra, Saini:2025ncj}. Here, rather than focusing on corrections to the orbital motion induced by the third body, we investigate how the presence of the tertiary affects the tidal interaction within the inner binary. Specifically, we examine the leading-order effects arising from a third body that is more massive than the inner binary. The influence of less massive tertiaries and other configurations will be explored in future work \cite{MartinBarandiaran:2023yav}.

When a BNS system evolves under mutual gravitational interaction, each star develops a multipolar deformation in response to the tidal field generated by its companion. These tidal interactions modify the equations of motion and consequently, imprint measurable signatures on the emitted gravitational waveform and its phase evolution. Tidal effects can be broadly classified into two categories: \textit{static tides} and \textit{dynamic tides}.  As the name suggests, a static tide refers to the quasi-stationary, non-zero multipole moment that a star acquires when immersed in a time-independent external tidal field. The static tidal response is typically characterized by the \textit{Love number}~\cite{Flanagan:2007ix}. 

The first GW observation of a BNS merger, GW170817, provided constraints on the neutron-star equation of state through measurements of this tidal Love number \cite{LIGOScientific:2017vwq, Nandi:2017rhy,  Landry:2018prl, Essick:2019ldf, Landry:2020vaw, Chatziioannou:2020pqz, Pacilio:2021jmq, Biswas:2021pvm, Biswas:2021paf}. On the other hand, dynamical tide becomes more important when oscillation modes of the neutron stars get excited and we encounter \textit{resonances} \cite{Flanagan:2007ix, Hinderer:2016eia, Steinhoff:2016rfi, Schmidt:2019wrl}. Using these resonances, constraints on f-modes of neutron stars have been found \cite{Pratten:2019sed}. In the current work, we will investigate both static and dynamic responses to the tidal field of a binary while it is under the influence of a third body in Newtonian gravity. Even though the Newtonian prescription is not accurate to explain strong-field regime, this will introduce the leading-order effect in a three-body system, and conveniently serve the purpose of the present work. With these results at hand, we will investigate the impact of such phenomena on GWs emitted by such systems. 

The rest of the paper is organized as follows. In Sec. \ref{sec:Formulation and Basic Equations}, we provide the basic foundation and derive the necessary equations to model the binary. This section is split into two parts, in one we discuss the effect without tide, and in the other we explore the effects with tide. The next section is devoted to studying the dephasing in the presence of the third body. Finally, we conclude the paper with a short discussion in Sec. \ref{sec:Discussion}.

\section{Formulation \& basic equations} \label{sec:Formulation and Basic Equations}
We introduce the basic setup and governing equations. Details are provided in Appendix \ref{Appendix A}, while the schematic description is presented in Fig.~ \ref{fig_01}. The smaller mass binary will be referred to as the \textit{inner binary}, and the binary consisting of the third body and the center of mass (COM) of the inner binary will be referred to as the \textit{outer binary}. We shall assume throughout that the three bodies and thus both binaries lie in the same plane. The position and masses of the $i$-th components are $\boldsymbol{x}_i$ and $m_i$ respectively, and the boldface indicates the vector nature of a quantity. The COM of the inner binary from the center of the third body of mass $m_3$ is $\boldsymbol{X}$. We will assume the third body's mass $m_3 \gg m_1+m_2 =M$.

\subsection{Without internal tide}
We consider a three-body system composed of a BNS and a distant massive object. The outer binary, constituting the COM of the inner binary, follows a circular path around the massive third body. We will focus only on the equatorial and circular solution for the COM of two bodies, i.e., $X^1 = \bar{X}\cos{\left(\Omega t \right)}$ and $X^2 = \bar{X}\sin{\left(\Omega t\right)}$, where $\Omega^2=m_3/\bar{X}^3$. The governing equation for the separation vector of the inner binary can be written as (see Eq.(\ref{eq: ri with epsilon})),
\begin{equation}\label{eq: ri with w^2}
    \dfrac{d^2 r^i}{dt^2} = -\dfrac{ M r^i}{r^3} - \epsilon K_{ij} r^j.
\end{equation}

\begin{figure}[t]  
\includegraphics[scale=.5]{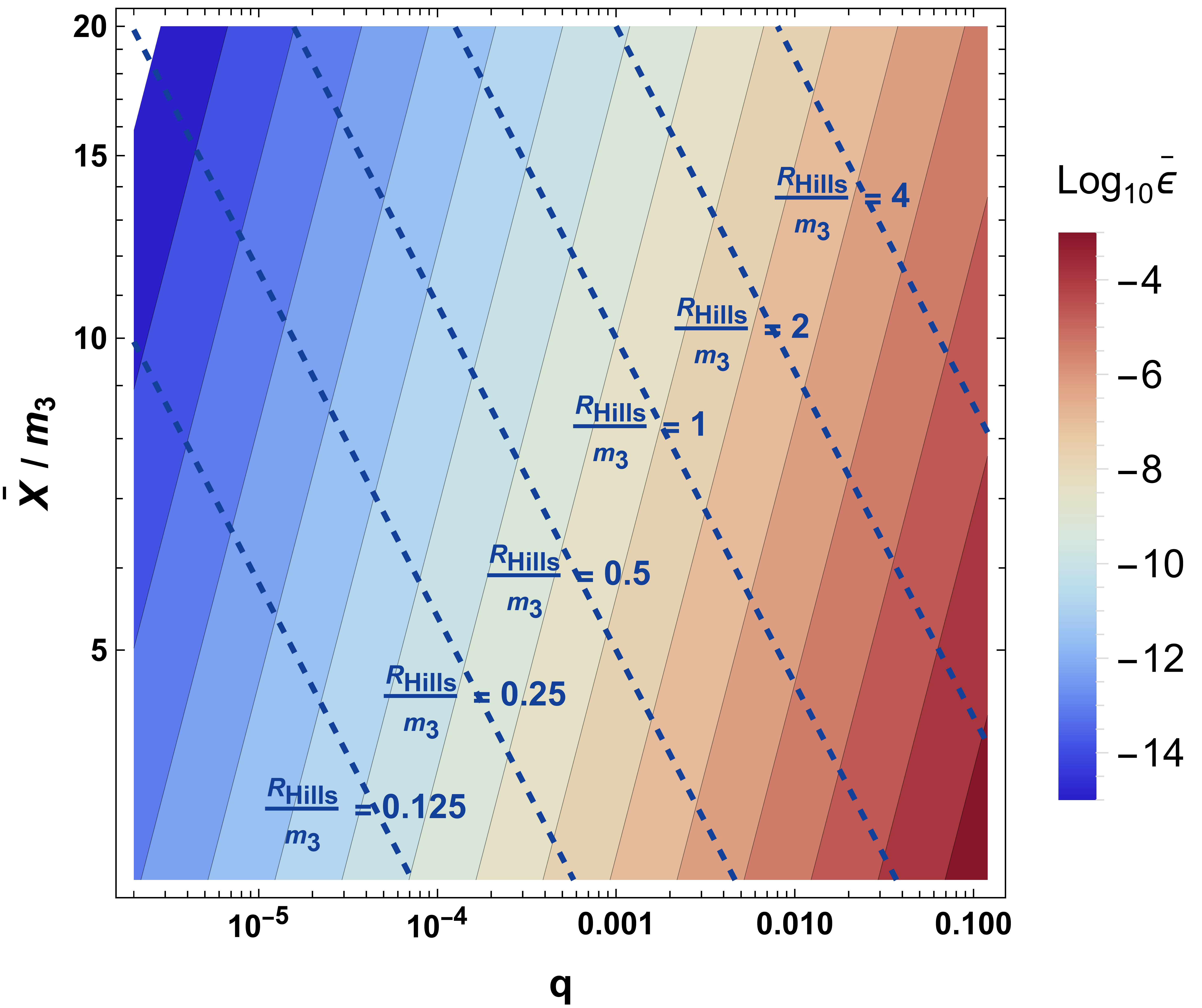}
    \caption{The figure shows the possible values of $\bar{\epsilon}$ for different binary configurations. On the x-axis, we show the mass ratio of the outer binary consisting of the third body and the inner binary, $q=M/m_3$, varying from $2\times10^{-6}$ to $0.12$. On the y-axis, we express the distance between the center of the third body and the COM of the inner binary, $\bar{X}$, in terms of the third body mass $m_3$. The figure also shows contour lines for constant Hills radius, $R_{\text{Hills}} \sim q^{1/3} \bar X$, expressed in units of $m_3$. A three-body system with a given $q$ and $\bar{X}$ will be in a tidally stable configuration, i.e., the binary will not be tidally disrupted by the third body, according to the Hills criterion, if the Keplerian radius of the binary satisfies $r_0 < R_\text{Hills}$.}
    \label{fig: epsilon bar}
\end{figure}

The expression of $K_{ij}$ depends only on $\Omega$ and is provided in Eq.(\ref{eq: Kij expression}). Note that for the equatorial plane, we can choose $i=3$ and the RHS of Eq.(\ref{eq: ri with w^2}) vanishes. Hence, if the initial configuration is such that $\Dot{r}^{i=3}=r^{i=3}=0$, it remains there. We define $\bar{\epsilon}=M^2\Omega^2 =M^2\epsilon$, and assume, $ \bar{\epsilon} \ll 1$ and $m_3 \gg M$. Therefore, $\bar{\epsilon}$ can be treated as a perturbative parameter. In Fig.~\ref{fig: epsilon bar} we show its range of values across different binary types, from b-EMRIs to b-IMRIs. The horizontal axis represents the mass ratio of the outer binary, defined as the ratio between the third body and the inner binary, $q = \,M/m_3$. We vary $q$ from $2\times10^{-6}$ to $0.12$. The vertical axis denotes the distance between the center of the third body and the COM of the inner binary, $\bar{X}$, expressed in units of the third body's mass $m_3$. Thus, the vertical axis measures how far the third body lies relative to its size, while the horizontal axis characterizes the disparity in mass between the inner and outer binaries.

The figure clearly demonstrates that $\bar{\epsilon} \ll 1$. The largest value it can attain is $\mathcal{O}(10^{-4})$. However, reaching such values requires the inner binary to be extremely close to the third body, effectively within the innermost stable circular orbit (ISCO) of the outer binary. In such a configuration, the inner binary would likely be tidally disrupted due to the Hills stability, $r_0 < R_\mathrm{Hills} \sim \left(\tfrac{m_1 + m_2}{m_3}\right)^{1/3}\bar{X}$, or $\omega > \Omega$ ~\cite{Santos:2025ass, Hills:1988cvl, Addison:2015bpa, Suzuki:2020, Souami:2020, Yin:2024nyz} (also see \cite{Camilloni:2023xvf, Cocco:2025udb}). This condition suggests that for the inner binary to survive tidal disruption, it must remain sufficiently far from the third body.

For simplicity, from here on we will use $r^i=(x,y,z)$. In the absence of internal tidal forcing terms, we take the ansatz,
\begin{eqnarray}
& & x = r(t) \cos\varphi, \quad y = r(t) \sin\varphi, \nonumber \\
& & r(t) = r_0 (1 + \bar{\epsilon } g(t))  , \quad \varphi=\omega t + \bar{\epsilon } f(t). 
\label{eq:xyr}
\end{eqnarray}
Here $\omega$ is the natural orbital frequency derived from Kepler's law, i.e., $\omega=\big({ M}/{r_0^3}\big)^{1/2}$. By writing it as above, we intend to condense the third-body effects in polar coordinates as a time-varying phase and radius. The equations satisfied by $g(t)$ and $f(t)$ are as follows: 
\begin{eqnarray}
  && 1 + 3 \cos(2t\omega) + 6 M^2 \omega^2 g(t) + 4 M^2 \omega \dot{f}(t)- 2 M^2\ddot{g}(t) = 0, \nonumber \\
  && 3 \sin(2t\omega) + 4 M^2 \omega \dot{g}(t) + 2 M^2 \ddot{f}(t) = 0
\end{eqnarray}

Solving the above equations, we find $f(t)$ and $g(t)$ to be:
\begin{eqnarray}
    f(t) = \frac{11 \sin (2 \omega  t)-8 \omega  t}{8 \omega ^2 M^2}, \nonumber \\
    g(t) = \frac{ 1-2 \cos (2 \omega  t)}{2 \omega ^2 M^2}\label{eq: no tide eom}.
\end{eqnarray}

\begin{figure*}
\centering
\includegraphics[width=86mm]{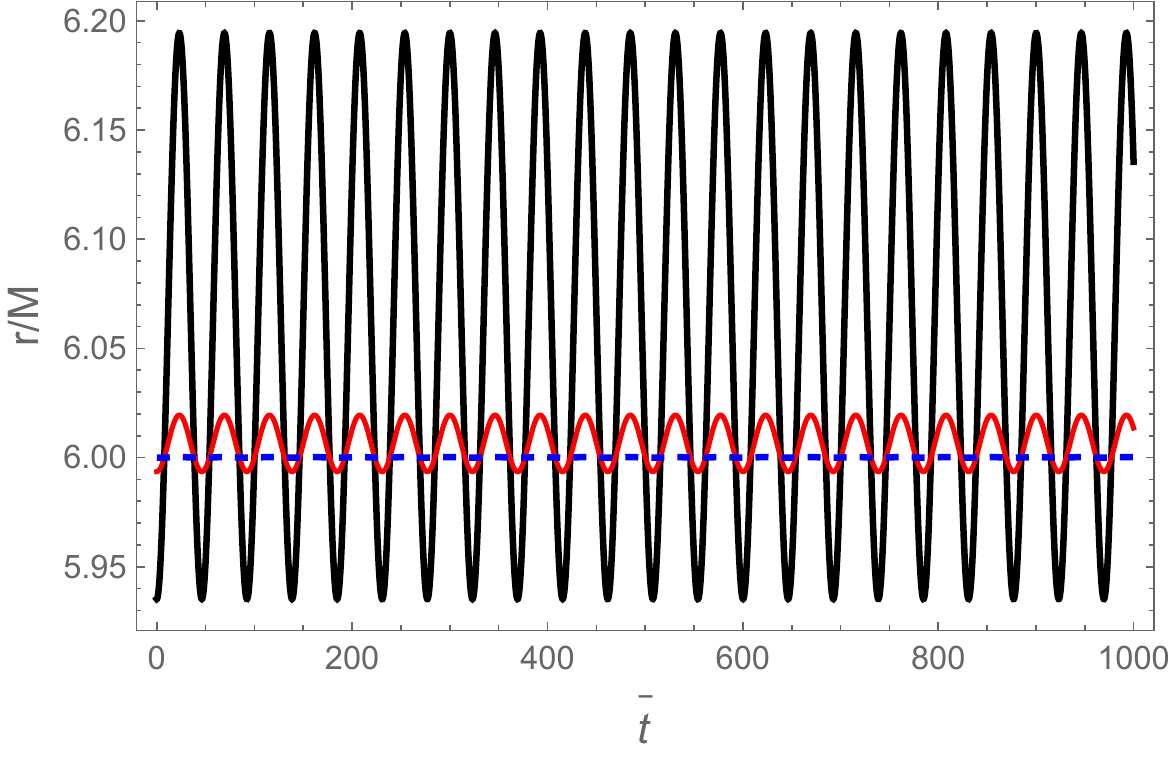}
\includegraphics[width=85mm]{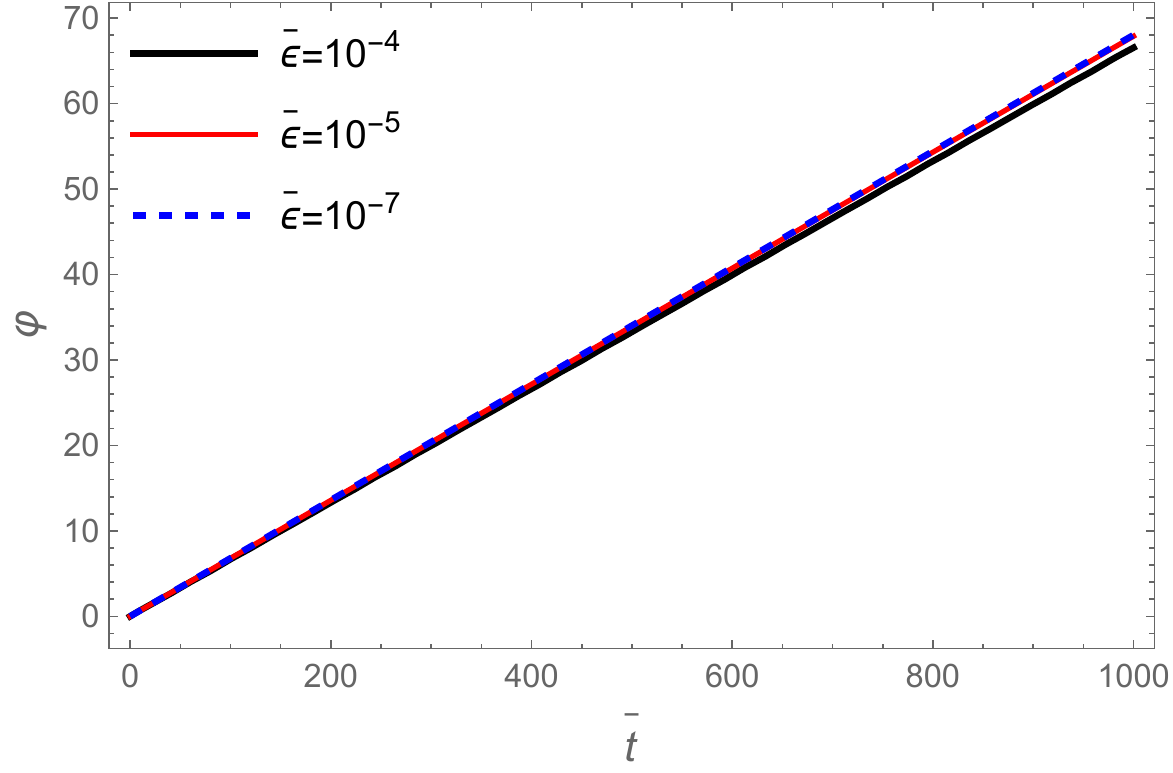}
\includegraphics[width=86mm]{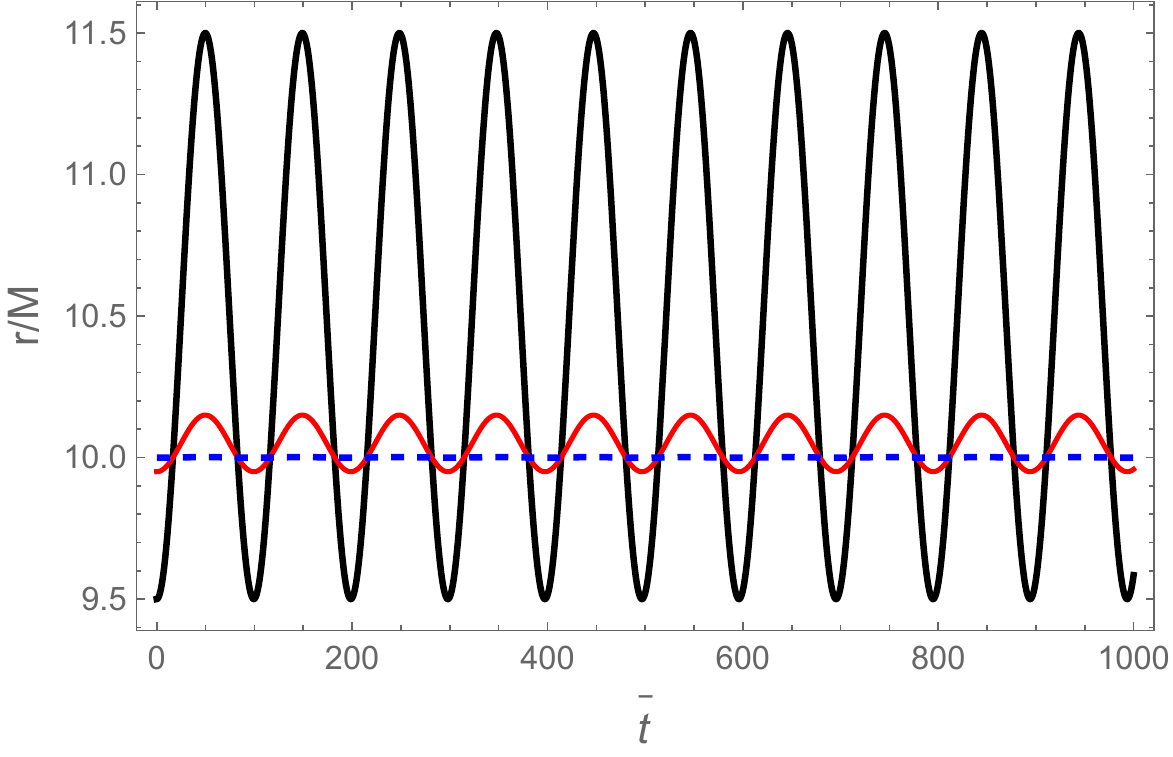}
\includegraphics[width=85mm]{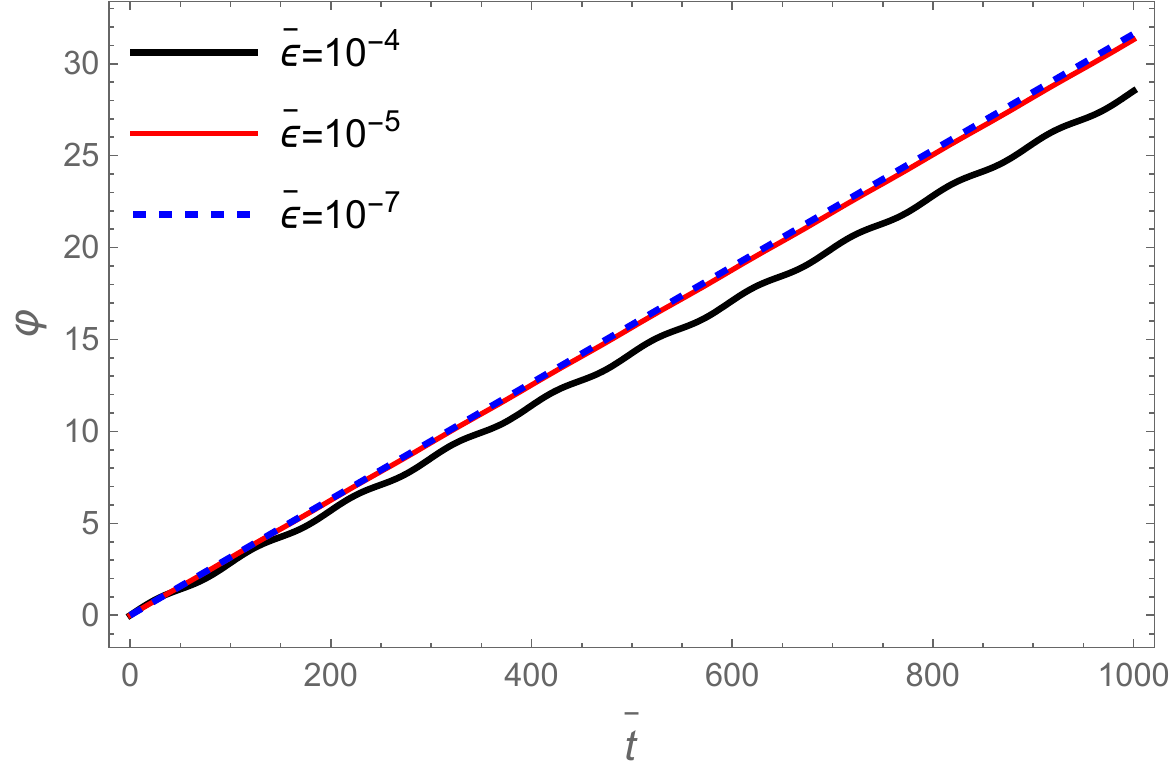}
\caption{In the above figure, we show how the presence of a third body can affect the orbital motions. In the left column, we plot $r(t)$ and in the right column, we plot $\varphi(t)$. Both are plotted against $\bar{t} = t/M$. In the first row, we consider $r_0=6M$, and in the second row, it is $10M$. The third body introduces an oscillation around $r_0$. It shows that the more compact a binary is, the less the absolute radial deviation from $r_0$. In the right column, the corresponding phase evolution is shown. Radiation reaction has not been taken into account while generating these plots.}   
\label{fig: eom no tide}
\end{figure*}

By using the above equations and employing Eq.(\ref{eq:xyr}), we can study how the radius and phase change for an orbit that experiences a third-body perturbation. This is demonstrated in Fig.\ref{fig: eom no tide}. In the left column, we show the solution of the radius for $r_0=6\textit{M}$ and $r_0= 10\textit{M}$ respectively. Three different values of $\bar{\epsilon}$ are chosen to demonstrate its impact. Due to the third body, the radial position oscillates about the reference radius $r_0$. With an increased $r_0$, the absolute deviation increases. This indicates that more compact binaries are less susceptible to third-body perturbations. This behavior is consistent with the Hills stability criterion. For a given $\bar{\epsilon}$, i.e., the third-body mass and position, a loosely held body is easier to disrupt tidally. In the right column, we show the phase of the motion. With an increased third-body tidal field, the dephasing increases. Similarly, the differences grow with time.

\subsection{With internal tide}
To introduce the tidal interaction in the inner binary, we follow the prescription of Ref.\cite{Flanagan:2007ix} and modify it appropriately. The effective action for the inner binary is,
\begin{equation}
    \begin{split}
        S = &\int dt \bigg[\frac{1}{2}\mu \dot r(t)^2+\frac{1}{2} \mu r(t)^2 \dot \phi(t)^2+\frac{M \mu}{r(t)}\bigg]\\
        &- \Bigg\{- \sum_n \int dt \frac{1}{4\lambda_{1,n} \omega_n^2} \bigg[\dot{Q}^{(1,n)}_{ij} \dot{Q}^{(1,n)}_{ij} - \omega_n^2 Q^{(1,n)}_{ij} Q^{(1,n)}_{ij}\bigg]\\
        &+\frac{1}{2}\int dt \, Q^{(1)}_{ij} \,\mathcal{E}_{ij}  +1\leftrightarrow2 \,\Bigg\}- \int dt \frac{\mu\epsilon }{2}K_{ij}r^i r^j,
    \end{split}
\end{equation}
where $M$ and $\mu$ are the total and reduced mass of the inner binary system; $\mathcal{E}_{ij}=\mathcal{E}^{(2)}_{ij}+\mathcal{E}^{(3)}_{ij}$ is the tidal field produced by the second body, $m_2$, acting on the first body, $m_1$, along with the third body's tidal field $\mathcal{E}^{(3)}_{ij}$; and $\omega_n$ is the oscillation frequency of the $n^\mathrm{th}$ oscillation mode. $Q^{(1,n)}_{ij}$ and $\lambda_{1,n}$ are the contributions to the total induced quadrupole moment ($Q^{(1)}_{ij}$) and tidal deformability ($\lambda_1$), of the first body for $l = 2$ and the $n^{\mathrm{th}}$ mode respectively. Thus $\lambda_1 \equiv \sum_n \lambda_{1,n}$ and $Q^{(1)}_{ij} \equiv \sum_n Q^{(1,n)}_{ij}$. The summation in the action thus corresponds to a sum over all $n$ oscillation modes. We drop the superscript `1' in $Q^{(1)}_{ij}$ and $Q^{(1,n)}_{ij}$ for brevity.

This action can be understood in terms of the energetics of the system as follows. The first bracket is $T-V$ for the inner binary in the COM frame of the inner binary, with $T$ and $V$ respectively being orbital kinetic energy and potential energy. The second line represents the action, $T_\mathrm{int} - V_\mathrm{int}$, of the first body's tidally induced quadrupole moment. The first term in the third line is the contribution arising from the interaction between the quadrupolar moment of the first body and the external tidal field, similar to Ref.~\cite{Flanagan:1997kp}. The final term is the potential energy from the interaction of the inner binary with the tidal field of the massive third body. For the total contribution of both bodies, a similar contribution should be added for the second body by interchanging the labels, represented as  $1\leftrightarrow2$. For simplicity, we compute only the excitation of star 1; the signals from the two stars add in the phase similar to the approach in \cite{Flanagan:2007ix}. Varying the action, we find,
\begin{align}
\label{eq:r tide 3body}
    \dfrac{d^2 r^i}{dt^2} + \dfrac{ M r^i}{r^3} =&  - \epsilon K_{ij} r^j +\frac{m_2}{2\mu}Q_{jk}\partial_i\partial_j \partial_k \frac{1}{r},\\
    \label{eq:QM eom 3 body}\Ddot{Q}^{(n)}_{ij} + \omega_n^2 Q^{(n)}_{ij} =& \lambda_{1,n}~\omega_n^2 \left(m_2\partial_i\partial_j\frac{1}{r} + m_3\partial_i\partial_j\frac{1}{\bar{X}} \right).
\end{align}
In the absence of the third body, i.e. $\epsilon\to 0$, Eq.~\ref{eq:r tide 3body} reduces to equations of Ref.~\cite{Flanagan:2007ix}. If tidal interaction is ignored, i.e. $Q_{ij}\to 0$, it recovers the equations of Ref.~\cite{Wardell2002} as discussed in Appendix~\ref{Appendix A}. Similarly, setting $m_3\to 0$ reduces Eq.~\ref{eq:QM eom 3 body} to equations in Ref.~\cite{Flanagan:2007ix} and in the limit of vanishing tide, the equation becomes irrelevant. The modification of Eq.(\ref{eq:QM eom 3 body}) can be understood directly from the effective action, which couples the induced quadrupole moment to the external tidal field. In this formulation, the right-hand side of Eq.(~\ref{eq:QM eom 3 body}) simply represents the tidal field acting as a driving force. In the presence of a third body, the total tidal field becomes,
\begin{equation}
    \mathcal{E}_{ij} = -m_2 \, \partial_i \partial_j \left(\frac{1}{r}\right) - m_3 \, \partial_i \partial_j \left(\frac{1}{\bar{X}}\right).
\end{equation}
This modification appears only in the interaction term of the action. Since $X^i$ is not a dynamical degree of freedom of the system, varying the action with respect to $r^i$ does not alter Eq.(\ref{eq:r tide 3body}), as the derivative $\partial_{r^i}$ acting on the third-body contribution vanishes. On the other hand, variation of the action with respect to $Q_{ij}$ introduces the total tidal field as the source term in the quadrupole evolution equation. The quadrupole oscillations respond to the total external tidal field.

With the equations at hand, we follow a similar method as in the section above. We use the ansatz,
\begin{eqnarray}
  && x=r(t) \cos(\omega t + \bar{\epsilon} f(t)+ \sum_n\bar{\lambda}_{1,n}\bar{\epsilon} F(t)), \nonumber \\ 
  && y = r(t) \sin(\omega t + \bar{\epsilon} f(t)+ \sum_n\bar{\lambda}_{1,n}\bar{\epsilon} F(t)), \nonumber \\ 
 && r(t) = r_0 +r_{\lambda}+ r_0 \bar{\epsilon} g(t)+ \sum_n\bar{\lambda}_{1,n} \bar{\epsilon}G(t),
\end{eqnarray}
where, $\bar{\omega}=M\omega$, $\bar{\omega}_n=M\omega_n$, $\bar{\lambda}_{1,n}M^5=\lambda_{1,n}$, $\bar{\epsilon} = M^2 \epsilon$ and
\begin{equation}
   r_{\lambda} =  \sum_n\frac{3 \bar{\omega}^{8/3} \bar{\lambda}_{1,n} M m_2 (\bar{\omega}^2-\bar{\omega}_n^2) }{m_1 \left(4
   \bar{\omega}^2-\bar{\omega}_n^2\right)},
\end{equation}
 which represents the correction in radius without any third-body effect \cite{Flanagan:2007ix}. While solving and taking the ansatz, we expand in the parameters $\bar{\epsilon}$ and $\bar{\lambda}_{1,n}$ while considering them as small and only consider terms up to and including the linear order and discarding terms from $\mathcal{O}(\bar{\epsilon}^2)$ and $\mathcal{O}(\bar{\lambda}_{1,n}^2)$ onwards. From the equations of motion, we obtain the differential equations for $F(t)$ and $G(t)$. Solving these two coupled differential equations, we obtain:
\begin{widetext}
    \begin{align}
    \label{eq: 3 body tide orbital solution}
   \frac{8 m_1}{3 \bar{\omega}^{4/3}} F(t) =& 12 \bar{\omega} \bar{t}
   \left(M-\frac{8 m_2 \left(4 \bar{\omega}^4-2 \bar{\omega}^2
   \bar{\omega}_n^2+\bar{\omega}_n^4\right)}{\left(\bar{\omega}_n^2-4 \bar{\omega}^2\right)^2}\right)+\sin (2 \bar{\omega}
   \bar{t}) \left(M+\frac{m_2 \left(-248 \bar{\omega}^4+207
   \bar{\omega}^2 \bar{\omega}_n^2-40 \bar{\omega}_n^4\right)}{64 \bar{\omega}^4-20 \bar{\omega}^2
   \bar{\omega}_n^2+\bar{\omega}_n^4}\right)\\
       \frac{4 m_1}{3 \bar{\omega}^{2/3} M } G(t) =& \cos (2 \bar{\omega} \bar{t})
   \left(M+\frac{8 m_2 \left(9 \bar{\omega}^4-11 \bar{\omega}^2 \bar{\omega}_n^2+2 \bar{\omega}_n^4\right)}{64
   \bar{\omega}^4-20 \bar{\omega}^2 \bar{\omega}_n^2+\bar{\omega}_n^4}\right)-3 M+\frac{24 m_2 \left(4
   \bar{\omega}^4-2 \bar{\omega}^2 \bar{\omega}_n^2+\bar{\omega}_n^4\right)}{\left(\bar{\omega}_n^2-4
   \bar{\omega}^2\right)^2},
\end{align}
\end{widetext}
where $r_0$ is given by $r_0 = ( M/\omega^2)^{1/3}$ and $\bar{t} = t/M$. The expressions for $Q_{ij}$ are provided in Appendix \ref{app: Q solutions}. These expressions will be used in the next section to compute energy, fluxes, and dephasing.

The dynamics presented above are derived within a controlled perturbative expansion in the small parameter $\bar\epsilon$, which quantifies the strength of the third-body perturbation on the inner binary and the tidal parameter $\bar{\lambda}_{1,n}$. All orbital quantities are consistently expanded to leading order in $\bar\epsilon$ and $\bar{\lambda}_{1,n}$; retaining Newtonian gravity, the $\mathcal{O}(\bar\epsilon)$ third-body contribution, and $\mathcal{O}(\bar{\lambda}_{1,n})$ contribution. Post-Newtonian (PN) corrections are not included in this work. For reference, spin-orbit effects enter at 1.5PN order, while spin-spin interactions enter at 2PN order. These terms are parametrically suppressed relative to the retained Newtonian and $\mathcal{O}(\bar\epsilon)$ contributions.

\section{Calculation of dephasing}
In this section, we use tidal response to study the GW emission in the current scenario. We use the tidal response in a full dynamical setting that is discussed in the last section. We start by defining the dynamic Love number as:
\begin{equation}\label{eq: lambda}
    \lambda^d = - \frac{\langle Q_{ij}\mathcal{E}_{ij}\rangle}{\langle \mathcal{E}_{ij}\mathcal{E}_{ij}\rangle},
\end{equation}
where $\langle\rangle$ represents the average of a quantity over an orbital time period. To compute the corresponding expression, we solve for $Q^{(n)}_{ij}$ from Eq.(\ref{eq:r tide 3body}) and then obtain the total quadrupole moment as $Q_{ij} = \sum_n Q^{(n)}_{ij}$. Averaging over the fast inner orbital time-scale, we obtain
\begin{equation}
    \frac{\lambda^d_{1,n}}{\lambda_{1,n} } = \frac{(1-x_n^2)}{(1-4x_n^2)}   -\frac{3
    \bar{\epsilon} \left\{M(1-4  x_n^2 ) +4
   m_2\right\}}{2 m_2 \left(1-4 x_n^2\right)^2 \bar{\omega}_n^2},
\end{equation}
where $x_n=\omega/\omega_n$. Note that in the static limit $(x_n \rightarrow 0)$, the $\lambda^d_{1,n}$ does not reduce to $\lambda_{1,n}$, rather it becomes
\begin{equation}
   \lim_{x_n \rightarrow 0}\lambda^d_{1,n}=\lambda_{1,n}\Big(1-\dfrac{3  \bar{\epsilon}(M+4m_2)}{2 m_2 \bar{\omega}_n^2}\Big).  
\end{equation}

Thus, the effective tidal Love number is shifted by the third body's gravitational field. However, although $\bar{\epsilon}$ may be static on the time scale of the inner binary, it varies on the orbital time scale of the outer binary. This contribution is also dynamic on longer timescales. Therefore, we recover the ``truly static response" to be $\lambda_{1,n}$.

In a conventional binary, two timescales govern the dynamics: the orbital timescale $t_{\rm orb}$ and the inspiral timescale $t_{\rm insp}$, with $t_{\rm insp} \gg t_{\rm orb}$ in the adiabatic regime. Orbital averaging therefore does not remove time dependence entirely. The corresponding evolution of the tidal response on the inspiral timescale is what we refer to as the dynamical Love number.

In a b-EMRI/IMRI-type three-body system, two additional scales appear: the outer orbital timescale $t_{\rm orb}^{\rm out}$ and its inspiral timescale $t_{\rm insp}^{\rm out}$, obeying the hierarchy $t_{\rm insp}^{\rm out} \gg t_{\rm orb}^{\rm out} > t_{\rm insp} \gg t_{\rm orb}$. While the static tidal response in an isolated binary is constant on $t_{\rm insp}$, in a three-body system it acquires slow evolution on $t_{\rm insp}^{\rm out}$ through its dependence on $\bar{\epsilon}$, which tracks the outer orbital frequency. Consequently, $\bar{\omega} \to 0$ alone does not recover the true static limit; one must also impose $\bar{\epsilon} \to 0$. Taking only $\bar{\omega} \to 0$ yields the static limit with respect to the inner binary, retaining a third-body–induced shift that is static on $t_{\rm insp}$ but not on $t_{\rm insp}^{\rm out}$.

An additional feature that emerges in the static limit $(x_n \rightarrow 0)$ is that there exists a regime, $3 \bar{\epsilon}(M+4m_2) \sim 2 m_2 \bar{\omega}_n^2$, where  $\lambda^d_{1,n}$ vanishes. Under the current approximation, this limit is not reachable, since $\bar{\omega}_n$ are the neutron star oscillation modes, and for $f-$modes they are $\sim \mathcal{O}(10^2)-\mathcal{O}{(10^3)}$~Hz. On the other hand, we have assumed the third body to be much larger than the BNS system; hence the frequency corresponding to $\bar{\epsilon}$ will be $<\mathcal{O}(\rm Hz)$. However, these could be relevant for $g-$modes. Alongside, this limit can be reached in a 3-body system comprising a third body with  $m_3 \sim M$. Although such systems have to be studied in a numerical setting, they may exhibit richer tidal interactions otherwise absent/suppressed in two-body configurations. As such, they also have the problem of being unstable and chaotic on fast timescales, whenever the hierarchy is relaxed, which would require careful investigation.

Before computing the radiated energy flux, it is useful to clarify the hierarchy of scales in the problem. The triple system contains two distinct orbital frequencies. The inner-binary frequency is $\omega$ and the outer-orbit frequency is $\Omega$, with $\Omega \ll \omega$ in the hierarchical regime considered here. Throughout this work, all quantities entering the tidal response, energy flux, and phase evolution are averaged over the fast inner orbital timescale. Consequently, the outer orbit is treated as a slowly varying tidal background, whose effect is encoded through the parameter $\bar{\epsilon}=M^2\Omega^2$. The GW phase computed below therefore corresponds to the high-frequency signal associated with the inner binary, i.e., the waveform component near frequency $2\omega$. In contrast, the outer orbit generates a separate low-frequency GW component near $2\Omega$, which lies in a different frequency band. The energy and flux computed below thus correspond to the orbit-averaged dynamics of the inner binary, and the resulting phase correction should be interpreted as an effective modification of the inner-binary waveform in the presence of the third-body tidal field. A complete treatment of dissipative energy exchange between the inner and outer orbital degrees of freedom, together with its impact on the full gravitational-wave signal of the triple system, is beyond the scope of the present work. Using the GW flux formula \cite{einstein1918gravitationswellen, 1964PhRv..136.1224P}
\begin{equation}
    \Dot{E}=-\dfrac{1}{5}\langle \dddot{Q}^{T(STF)}_{ij}\dddot{Q}^{T(STF)}_{ij}\rangle,
\end{equation}
where, $Q^{T}_{ij}= \sum_n Q^n_{ij} + \mu x_ix_j - \mu r^2 \delta_{ij}/3$, and $STF$ represents symmetric-trace-free, energy loss from the system can be computed. Similarly, using the action, the Hamiltonian of the system can be computed to find the energy to be,
\begin{eqnarray}
    E &=& -\dfrac{\mu v^{2}}{2} \left[1+ \sum_n \chi_{n}  g_2(x_n) \right], \\
        \dot{E} &=& -\dfrac{32 \mu^2 v^{10}}{5 M^2} \left[1+ \sum_n \chi_{n}  g_3(x_n) \right],
\end{eqnarray}
where $v = (M \omega)^{1/3}$, and $g_i$ is defined as,
\begin{equation}
    g_i = g_i^{0} + \frac{\bar{\epsilon}}{v^6} g_i^{\epsilon},
\end{equation}
and $\chi_n m_1 M^{5/3} =  m_2\lambda_{1,n}\omega^{10/3}$, with,
\begin{eqnarray}
    g_2^{0} &=&   -\frac{9 \left(4x_n^4-3
   x_n^2+1\right)}{(1-4x_n^2)^2}, \\
   g_2^{\epsilon} &=& \frac{9M}{m_2} + \frac{3 \left(400 x_n^6-312 x_n^4+177
   x_n^2-25\right) }{(1-4x_n^2)^3 }, \\
   g_3^{0} &=&   \frac{6 \left(M-2 m_2 x_n^2+2 m_2\right)}{m_2(1-4x_n^2)},   \\
   g_3^{\epsilon} &=& -\frac{3 M \left(7 + 12 x_n^2 - 96 x_n^4\right)}{m_2(1-4x_n^2)^2 } \nonumber \\ &-& \frac{18 m_2 \left(44x_n^4-31 x_n^2+11\right)}{m_2(1-4x_n^2)^2 }.
\end{eqnarray}

Using the relation \cite{Tichy:1999pv,Flanagan:2007ix}
\begin{eqnarray}
\dfrac{d^{2}\Psi}{d\omega^{2}} = \dfrac{2}{\dot{E}} \dfrac{dE}{d\omega},    
\end{eqnarray}
for the phase, $\Psi(f)$, of the Fourier transform of the GW signal, with $f = \omega/\pi$, and by following Ref.~\cite{Tichy:1999pv}, we derive the corresponding tidal phase correction to be,
\begin{equation}
    \delta \psi = -\frac{15m_2^2}{16\mu^2 M^5} \sum_n \lambda_{1,n} \int_{v_I}^v dv' v' (v^3 - v'^3) g_4(x_n'),
\end{equation}
where $v=(\pi M f)^{1/3}$, $v_I$ is the initial velocity related to the initial time and phase of the waveform, and
\begin{equation}
    g_4^{0} = \frac{2M}{m_2} \frac{1}{1-4x_n^2} + \frac{22-117 x_n^2 +348 x_{n}^4 -352 x_{n}^6}{(1-4 x_n^2)^3},
\end{equation}
\begin{equation}
    \begin{split}
        g_4^{\epsilon} =& \frac{ 2M \left(3 - 14 x_n^2 - 24 x_n^4\right)}{m_2 
   \left(1-4 x_n^2\right)^2} \\
   &+\frac{ \left(8000 x_n^8-9824 x_n^6+2820
   x_n^4-932
   x_n^2+125\right)}{ \left(1-4
   x_n^2\right)^4}.
    \end{split}
\end{equation}

In the static limit $x_n\rightarrow 0$, resulting in $g_4^{0}\rightarrow \frac{2M}{m_2} +22$, $g_4^{\epsilon} \rightarrow \frac{6M}{m_2}+125$ and $\lambda_1 \equiv \sum_n \lambda_{1,n}$. Hence, the phase reduces to,
\begin{equation}
\label{eq: dephasing 3 body static tide}
    \begin{split}
        \delta\psi =& -\frac{9 \lambda_{1} v^5 }{16M^4 \mu} \left(\frac{M}{m_1}+11 \frac{m_2}{m_1}\right) + \delta \psi_{\bar{\epsilon}}+ 1\leftrightarrow 2,\\
        \delta \psi_{\bar{\epsilon}}=  &-\frac{45 \bar{\epsilon} \lambda_{1} }{64M^4 \mu v} \left(6 \frac{M}{m_1}+125 \frac{m_2}{m_1}\right).
    \end{split}
\end{equation}

\begin{figure}[htp]
\centering
\includegraphics[scale=.4]{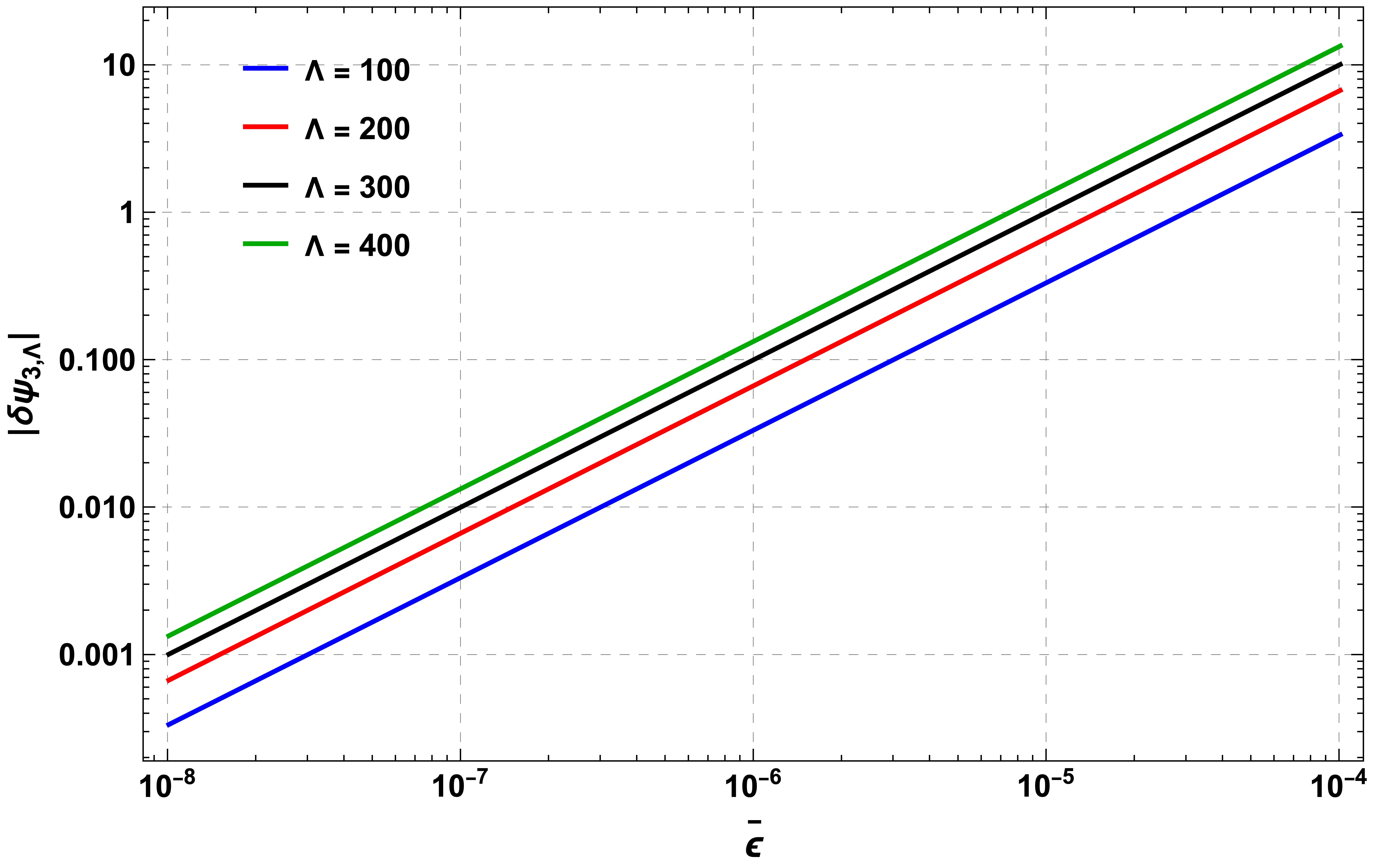}
    \caption{In the above figure, we show the accumulated dephasing due to the modification in the static tidal interaction due to the third body. The accumulation is computed using ($\delta \psi_{\bar{\epsilon}}+1\leftrightarrow2$) from Eq.(\ref{eq: dephasing 3 body static tide}) and integrating from 10 Hz to ISCO. Total mass of the equal-mass inner binary $M=2.8 M_{\odot}$. Different colors represent different tidal deformability. On the horizontal axis, the strength of the third body's field, $\bar{\epsilon}$, is shown. The accumulated dephasing reaches $\mathcal{O}(1)$ for $\bar{\epsilon}\sim10^{-5}$. However, in realistic binaries that can survive tidal disruption, with smaller $\bar{\epsilon}$, the accumulation is significantly lower. Systems with larger $\bar{\epsilon}$ are more likely to show a larger observable imprint of the modified tidal interaction.
    }
    \label{fig:dephasing}
\end{figure}

In Fig.~\ref{fig:dephasing} we plot the accumulated dephasing for the static tide due to the additional third body for an equal-mass BNS of $M=2.8M_{\odot}$ with different values of $\Lambda =\Lambda_i\equiv \lambda_i/m_i^5$. We use $\delta \Psi_{3,\Lambda} = 2\int_{10 {\rm Hz}}^{f_{\rm ISCO}} df f d^2 \delta \psi_{\bar{\epsilon}}/df^2$. For an equal-mass system, each component of the inner binary will contribute equally to the total dephasing, resulting in the factor 2 in the accumulated phase.

We use lowercase $\psi(f)$ to denote the frequency-domain gravitational-wave phase appearing in the waveform. Uppercase $\Psi$ denotes the accumulated phase shift across the detector frequency band, obtained by integrating the corresponding phase difference over the relevant frequency interval. The quantity $\delta \Psi_{3,\Lambda}$ represents the accumulated third-body correction to the tidal phase over the band.

On the horizontal axis, we keep the third body's field strength $\bar{\epsilon}$. We show the results for $\Lambda=100,\,200,\,300,\,400$, keeping the largeness of the effect consistent with the current constraints~\cite{LIGOScientific:2017vwq, Biswas:2021paf}. As expected, the dephasing increases with increasing $\bar{\epsilon}$ and $\Lambda$. Interestingly, even for $\bar{\epsilon}\sim \mathcal{O}(10^{-5})$ the accumulated dephasing can be $\mathcal{O}(1)$. For a larger value of $\bar{\epsilon}=10^{-4}$, the accumulation can reach over $\sim10$ radians for $\Lambda=400$, making the effect potentially observable provided the system survives tidal disruption. The impact can be understood by comparing with Fig.~\ref{fig: epsilon bar}. For $\bar{\epsilon}$ to reach $10^{-4}$, the third-body mass must be closer to the inner binary, i.e. $q\sim0.06-0.1$, where the present perturbative treatment becomes inapplicable and fully numerical approaches will be required \cite{Cardoso:2026ugm, Campanelli:2007ea}.

Furthermore, such values of $\bar{\epsilon}$ can be attained only for very compact outer orbits ($\bar{X}\sim6$), i.e. close to the ISCO of the primary. In this regime, the adiabatic approximation for the outer orbit may no longer be valid for low-mass primaries, as gravitational-radiation reaction can cause the orbital frequency $\Omega$ (and hence $\bar{\epsilon}$) to evolve appreciably over the duration of the observed signal from the inner binary (i.e. for BNS). This will invalidate the assumption that the third body provides a slowly varying tidal background. Consequently, the largest dephasings presented here should be regarded as illustrative of the maximum possible effect rather than quantitatively reliable predictions. For $q<10^{-4}$, $\bar{\epsilon}$ reaches very small values; hence, in b-EMRI systems the dephasing is not expected to be appreciable. In the range $0.05>q>10^{-3}$, $\bar{\epsilon}\sim10^{-5}-10^{-8}$, requiring SNR $\sim10-1000$ for the effect to be observable \cite{Flanagan:1997kp,Lindblom:2008cm}. Therefore, b-IMRI systems constitute the most promising regime where the present analysis is expected to be applicable yielding potentially observable dephasing.

However, in realistic binaries that are more likely to survive tidal disruption, with smaller $\bar{\epsilon}$, the accumulation is significantly lower. Among the different types of three-body systems considered, b-IMRI systems are more relevant for these kinds of effects to be observed.

\section{Discussion} \label{sec:Discussion}
We investigate how tidal interactions of a compact binary system are modified in the presence of an additional third body. To model the inner binary's tidal dynamics, we adopted the effective field theory framework introduced in Ref.~\cite{Flanagan:2007ix}. To model the inner binary's tidal dynamics, we adopted the framework introduced in Ref.~\cite{Flanagan:2007ix}. We then extended this approach to incorporate the effect of a third body on the equations of motion of the inner binary, following the methodology outlined in Ref.~\cite{Wardell2002}.

To modify the evolution equation of the quadrupole moments of the stars, we draw motivation from the action described in Eq.(\ref{eq:QM eom 3 body}). Since the interaction term couples the tidal field directly to the quadrupole moment, the tidal field naturally acts as a source term in the evolution equation of the quadrupole moment. Importantly, the response of a star to an external tidal field should depend on the total tidal field the star is immersed in. Thus, the quadrupole evolution should account not only for the field generated by the binary companion but also for the tidal influence of the third body. Consequently, the tidal field in the action can be interpreted as the total tidal environment of the system, rather than just that of the inner binary. This observation is central to the present analysis. This modifies the evolution equation of the quadrupole moments, introducing a forcing term arising from the third body's field.

With these modifications, we obtained explicit solutions for both the orbital configuration and the quadrupole moment, summarized in Eq.(\ref{eq: 3 body tide orbital solution}) and App. \ref{app: Q solutions}. From these solutions, we calculated the total energy of the system and energy loss from it using the standard flux formula with orbital averaging. The energy balance then allowed us to derive an integral expression for the GW dephasing under the stationary-phase approximation. In the static limit, this integration can be carried out exactly, and we report the closed-form result.  

Using this dephasing expression, we examined the impact of the third body on tidal interactions. In optimistic scenarios, the accumulated dephasing can exceed $\mathcal{O}(1)$. For more astrophysically realistic configurations, where the inner binary can survive tidal disruption, the dephasing is significantly smaller. In b-EMRI systems, the accumulated dephasing remains small unless the center of mass of the inner binary passes extremely close to the third body. In contrast, b-IMRI systems appear to be more promising candidates, where third-body tidal effects can leave observable imprints. Up to the inner-binary ISCO, such systems may accumulate a dephasing in the range $\sim 0.01-1$, and can be potentially detectable by third-generation gravitational-wave observatories \cite{Evans:2021gyd, Abac:2025saz}. A detailed study of formation, survival, and evolution is left for future work.

In the presence of a third body, the orbital configuration, with or without tidal interaction, of the inner binary deviates from a circular orbit. For such a binary, energy flux alone may not be sufficient to describe the GW dephasing. As an initial investigation of this problem, we do not include the effects of angular momentum loss and the additional dephasing it may induce. In contrast to the circular-orbit case, where this is not required, these effects may become relevant here and will be investigated in future work.

An important limitation of the present analysis concerns the treatment of dissipation in a hierarchical triple system. The full GW signal contains contributions associated with both the inner orbital frequency $\omega$ and the outer orbital frequency $\Omega$. Because $\Omega \ll \omega$, these contributions occupy widely separated frequency bands. In this work, we focus on the high-frequency sector associated with the inner binary and derive the corresponding modification to its tidal response and phase evolution. The outer orbit is treated as a slowly varying tidal environment, whose effect is incorporated through the parameter $\bar{\epsilon}$. The present analysis assumes that the primary provides a slowly varying tidal background. A complete analysis of energy exchange between the inner and outer orbital sectors, together with its impact on the full gravitational-wave signal of the triple system along with a self-consistent treatment of the coupled evolution of the inner and outer binaries, including radiation reaction of the outer orbit, is left for future work.

\begin{acknowledgments}
S.D. acknowledges financial support from MUR, PNRR - Missione 4 - Componente 2 - Investimento 1.2 - finanziato dall'Unione europea - NextGenerationEU (cod. id.: $SOE2024\_0000167$, CUP:D13C25000660001). S.M. is thankful to the Inspire Faculty Grant (DST/INSPIRE/04/2020/001332) from DST, Govt. of India, Prime Minister Early Career Research Grant (ANRF/ECRG/2024/004108/PMS) by ANRF, Govt. of India, and the New Faculty Seed Grant (NFSG/PIL/2023/P3794) provided by BITS Pilani (Pilani), India, for financial support. He (S.M.) is also grateful to the Visiting Associateship Program at IUCAA, Pune, for academic visits where a part of this work was carried out. MK is grateful for the JRF position at RRI, Bengaluru, where a part of the work was carried out.
\end{acknowledgments}

\appendix \label{appendix}
\section{Equations of motion of the three-body system} \label{Appendix A}
Throughout this Appendix we employ the same perturbative ordering as in Sec.~\ref{sec:Formulation and Basic Equations}, expanding all quantities in powers of the small parameter $\bar\epsilon \ll 1$ and retaining only leading-order terms. Higher-order relativistic (post-Newtonian) corrections, including spin-orbit and spin-spin effects, are systematically neglected, as they scale with additional powers of $v/c$ relative to the Newtonian dynamics.

To obtain the equations of motion of the three-body system in the absence of tides, we follow the basic structure used in Ref. \cite{Wardell2002}. Given that the position vector of each body (treated here as a point mass) is given by $\boldsymbol{x_1}$, $\boldsymbol{x_2}$, and $\boldsymbol{x_3}$; and $x_j^i$ is denoted as the $i^\mathrm{th}$ component of the $j^\mathrm{th}$ body's position, we arrive at:
\begin{equation}
    m_1 \dfrac{d^2 x_1^i}{dt^2} + \dfrac{ m_1 m_2 (x_1^i - x_2^i)}{|\boldsymbol{x_1} - \boldsymbol{x_2}|^3} = - \dfrac{ m_1 m_3 (x_1^i - x_3^i)}{|\boldsymbol{x_1} - \boldsymbol{x_3}|^3},
\end{equation}
\begin{equation}
    m_2 \dfrac{d^2 x_2^i}{dt^2} + \dfrac{ m_1 m_2 (x_2^i - x_1^i)}{|\boldsymbol{x_1} - \boldsymbol{x_2}|^3} = - \dfrac{ m_2 m_3 (x_2^i - x_3^i)}{|\boldsymbol{x_2} - \boldsymbol{x_3}|^3},
\end{equation}
\begin{equation}
\label{eq: m3 motion}
    m_3 \dfrac{d^2 x_3^i}{dt^2} + \dfrac{ m_1 m_3 (x_3^i - x_1^i)}{|\boldsymbol{x_1} - \boldsymbol{x_3}|^3} = - \dfrac{ m_2 m_3 (x_3^i - x_2^i)}{|\boldsymbol{x_3} - \boldsymbol{x_2}|^3}.
\end{equation}

In the present work, we consider $m_3 >> m_1 + m_2$, and $x_3^i - x_{(1,2)}^i\geq 6 m_3 $, which ensures that the inner binary remains outside the ISCO of the third body. Therefore, Eq.(\ref{eq: m3 motion}) simplifies to $\dfrac{d^2 x_3^i}{dt^2}\sim 0$. Assuming the third body has zero velocity at the beginning, $x_3^i$ becomes constant. Hence, the coordinate origin can be shifted to the position of $m_3$, without losing any generality. Hence, we fix $\boldsymbol{x_3} = 0$, and in this frame,
\begin{equation}\label{eq: 1com}
    m_1 \dfrac{d^2 x_1^i}{dt^2} = - \dfrac{ m_1 m_3 x_1^i}{|\boldsymbol{x_1}|^3} + \dfrac{ m_1 m_2 (x_2^i - x_1^i)}{|\boldsymbol{x_2} - \boldsymbol{x_1}|^3},
\end{equation}
\begin{equation}\label{eq: 2com}
    m_2 \dfrac{d^2 x_2^i}{dt^2} = - \dfrac{ m_2 m_3 x_2^i}{|\boldsymbol{x_2}|^3} - \dfrac{ m_1 m_2 (x_2^i - x_1^i)}{|\boldsymbol{x_2} - \boldsymbol{x_1}|^3}.
\end{equation}

We now make a change of coordinates as follows:
\begin{eqnarray}\label{eq: change of coords}
    r^i &=& x_2^i - x_1^i,\\
    X^i &=& \dfrac{m_1 x_1^i + m_2 x_2^i}{M},
\end{eqnarray}
where $r^i$ is the separation between the two masses, $X^i$ is the separation between the centre of mass of the binary and the third body, and $M = m_1 + m_2$. Using Eq.(\ref{eq: 1com}), Eq.(\ref{eq: 2com}), and Eq.(\ref{eq: change of coords}), the equation of relative motion of the binary system becomes:
\begin{equation}\label{eq: ri}
    \dfrac{d^2 r^i}{dt^2} = - \dfrac{ M r^i}{|\boldsymbol{r}|^3} -  m_3 \left( \dfrac{x_2^i}{|\boldsymbol{x_2}|^3} - \dfrac{x_1^i}{|\boldsymbol{x_1}|^3} \right)
\end{equation}

The equation of motion of the COM of the binary becomes:
\begin{equation}\label{eq: Xi}
    \dfrac{d^2 X^i}{dt^2} = - \dfrac{ m_3}{M} \left( \dfrac{m_1 x_1^i}{|\boldsymbol{x_1}|^3} + \dfrac{m_2 x_2^i}{|\boldsymbol{x_2}|^3} \right)
\end{equation}

We now invert Eq.(\ref{eq: change of coords}) to obtain $x_1^i$ and $x_2^i$ in terms of $r^i$ and $X^i$:
\begin{eqnarray}
    x_1^i = X^i - \gamma_2 r^i, \\
    x_2^i = X^i + \gamma_1 r^i,
\end{eqnarray} \\
where $\gamma_2 = m_2/M$ and $\gamma_1 = m_1/M$. Thus, we can get Eq.(\ref{eq: ri}) and Eq.(\ref{eq: Xi}) purely in terms of $X^i$ and $r^i$ as follows:
\begin{eqnarray}
    \label{eq: ri with Kij}
    \dfrac{d^2 r^i}{dt^2} &=& - \dfrac{ M}{r^3} r^i - \dfrac{ m_3}{\bar{X}^3} K_{ij} r^j,\\
    \label{eq: Xi eqn}
    \dfrac{d^2 X^i}{dt^2} &=& - \dfrac{ m_3}{\bar{X}^3} X^i,
\end{eqnarray}
where $K_{ij} = \left( \delta_{ij} - \tfrac{3 X^i X^j}{\bar{X}^2} \right)$ is the reduced tidal matrix and $\bar{X}^2= \delta_{ij}X^i X^j$. For the present work, we will only focus on the equatorial circular solution of the COM of the two bodies: $X^1 = \bar{X}\cos{\left(\Omega t \right)}$ and $X^2 = \bar{X}\sin{\left(\Omega t\right)}$, where $\Omega = \sqrt{\frac{ m_3}{\bar{X}^3}}$. Therefore, Eq.(\ref{eq: ri with Kij}) will now become:
\begin{equation}\label{eq: ri with w^2 in appendix}
    \dfrac{d^2 r^i}{dt^2} = - \dfrac{ M}{r^3} r^i - \Omega^2 K_{ij} r^j
\end{equation}

Using the values of $X^1$ and $X^2$, we obtain the $K_{ij}$ matrix to be:
\begin{equation}
\label{eq: Kij expression}
    K_{ij} = \begin{bmatrix}
    1-3 \cos^2{ \Omega t} & -3 \cos{ \Omega t}\sin{ \Omega t} & 0\\
    -3 \cos{ \Omega t}\sin{ \Omega t} & 1-3 \sin^2{ \Omega t} & 0\\
    0 & 0 & 1
\end{bmatrix}
\end{equation}

The third-body perturbation term is assumed to be very small ($M^2\Omega^2 << 1$), so it proves useful to replace it with a free parameter $\bar{\epsilon}=M^2\epsilon =M^2\Omega^2<< 1$ in Eq.(\ref{eq: ri with w^2}), and we finally get
\begin{equation} \label{eq: ri with epsilon}
    \dfrac{d^2 r^i}{dt^2} = -\dfrac{ M r^i}{r^3} - \epsilon K_{ij} r^j.
\end{equation}

This equation is our starting point in Eq.(\ref{eq: ri with w^2}).

\section{Dynamic solution of $Q_{ij}$} \label{app: Q solutions}
In this section, we demonstrate the solution of Eq.(\ref{eq:QM eom 3 body}). To compute the full tidal response and modification in the energy, fluxes, and phasing, as well as the solution for Eq.(\ref{eq:r tide 3body}), it is necessary to obtain the solution for Eq.(\ref{eq:QM eom 3 body}). The $\mathcal{O}(\bar{\lambda})$ solution is already provided in Ref. \cite{Flanagan:2007ix}. Here, we will only provide the solution for $\mathcal{O}(\bar{\epsilon }\bar{\lambda})$:
\begin{widetext}
    \begin{align}
    Q^{(n)}_{11} =& \bar{\lambda}_{1,n} \bar{\epsilon}\left(-\frac{3 \bar{\omega} M^2 m_2
   \bar{t} \bar{\omega}_n^2 \sin (2 \bar{\omega} \bar{t})}{4
   \bar{\omega}^2-\bar{\omega}_n^2}-\frac{3 M^2 m_2
   \bar{\omega}_n^2 \left(12 \bar{\omega}^2+\bar{\omega}_n^2\right) \cos (2 \bar{\omega} \bar{t})}{4 \left(\bar{\omega}_n^2-4 \bar{\omega}^2\right)^2}+\frac{69 M^2 m_2 \bar{\omega}_n^2 \cos (4 \bar{\omega} \bar{t})}{16 \left(\bar{\omega}_n^2-16 \bar{\omega}^2\right)}+2 M^3-\frac{9 M^2 m_2}{16}\right) \\
   Q^{(n)}_{22} =& \bar{\lambda}_{1,n} \bar{\epsilon} \left(\frac{3 \bar{\omega} M^2
   m_2 \bar{t} \bar{\omega}_n^2 \sin (2 \bar{\omega}
   \bar{t})}{4 \bar{\omega}^2-\bar{\omega}_n^2}+\frac{3 M^2
   m_2 \bar{\omega}_n^2 \left(5 \bar{\omega}_n^2-4 \bar{\omega}^2\right) \cos (2 \bar{\omega} \bar{t})}{4 \left(\bar{\omega}_n^2-4 \bar{\omega}^2\right)^2}+\frac{69 M^2 m_2 \bar{\omega}_n^2 \cos (4 \bar{\omega} \bar{t})}{256 \bar{\omega}^2-16
   \bar{\omega}_n^2}-\frac{1}{16} M^2 (16 M+15 m_2)\right) \\
   Q^{(n)}_{12} =& \bar{\lambda}_{1,n} \bar{\epsilon} \left(\cos (2 \bar{\omega}
   \bar{t}) \left(\frac{3 \bar{\omega} M^2 m_2 \bar{t}
   \bar{\omega}_n^2}{4 \bar{\omega}^2-\bar{\omega}_n^2}+\frac{69 M^2 m_2 \bar{\omega}_n^2 \sin (2 \bar{\omega} \bar{t})}{8 \left(\bar{\omega}_n^2-16 \bar{\omega}^2\right)}\right)-\frac{3 M^2 m_2 \bar{\omega}_n^2 \left(4
   \bar{\omega}^2+3 \bar{\omega}_n^2\right) \sin (2 \bar{\omega} \bar{t})}{4 \left(\bar{\omega}_n^2-4 \bar{\omega}^2\right)^2}\right) \\
   Q^{(n)}_{33} =& \bar{\lambda}_{1,n} \bar{\epsilon} \left(\frac{1}{2} M^2 (3 m_2-2
   M)-\frac{3 M^2 m_2 \bar{\omega}_n^2 \cos (2 \bar{\omega}
   \bar{t})}{\bar{\omega}_n^2-4 \bar{\omega}^2}\right)
\end{align}
\end{widetext}

The above expression has been used along with the solution of Eq.(\ref{eq:r tide 3body}) to find the energy, fluxes, and the corresponding dephasing discussed in the main text of the paper.

\bibliography{references}

\end{document}